\PassOptionsToPackage{
  hidelinks, 
  hypertexnames=false,
}{hyperref}

\documentclass[runningheads]{llncs}

\usepackage{tikz}
\usepackage{pgfplots}
\pgfplotsset{compat=1.18}
\usetikzlibrary {arrows.meta,positioning}
\tikzset{>=stealth}
\usepackage{graphicx}
\graphicspath{{Figs/}}
\usepackage{hyperref}

\usepackage[nameinlink]{cleveref}

\usepackage{enumitem}
\usepackage{xspace}

\usepackage{caption}
\usepackage{subcaption}

\usepackage[backend=biber, style=lncs]{biblatex}
\usepackage{adjustbox}
\usepackage{tabularx}
\usepackage{multirow}
\usepackage{booktabs}
\usepackage[table]{xcolor}
\usepackage{listings}
\usepackage{ragged2e} 
\usepackage{makecell}
\usepackage{cancel}

\newcommand{\qwencoderB}{blue}
\newcommand{\qwencoderF}{\qwencoderB!40}
\newcommand{\qwenB}{orange}
\newcommand{\qwenF}{\qwenB!60}
\newcommand{\llamaB}{purple}
\newcommand{\llamaF}{\llamaB!60}
\newcommand{\chatgptB}{brown}
\newcommand{\chatgptF}{\chatgptB!60}
\newcommand{\deepseekB}{red!50!black}
\newcommand{\deepseekF}{\deepseekB!60}
\newcommand{\codegptB}{olive}
\newcommand{\codegptF}{\codegptB!60}
\newcommand{\codegenB}{green!40!black}
\newcommand{\codegenF}{\codegenB!50}
\newcommand{\codetfiveB}{magenta}
\newcommand{\codetfiveF}{\codetfiveB!60}

\title{RedShell: A Generative AI-Based Approach to Ethical Hacking}
\titlerunning{RedShell}
\author{%
    Ricardo Bessa%
    \inst{1,2}~%
\and
    João Trindade%
    \inst{2}%
\and
    Rui Claro%
    \inst{2}%
\and
    João M. Lourenço%
    \inst{1}~%
}
\authorrunning{R.\ Bessa et al.}
\institute{%
    NOVA University Lisbon — FCT \& NOVA LINCS, Portugal \\
    \email{r.bessa@campus.fct.unl.pt} \qquad \email{joao.lourenco@fct.unl.pt}
\and 
    Layer8 - Shield Domain SA, Portugal \\
    \email{joao.trindade@layer8.pt} \qquad \email{rui.claro@layer8.pt}
}

\begin{document}

\maketitle

\begin{abstract}
The application of Machine Learning techniques in code generation is now a common practice for most developers. Tools such as ChatGPT from OpenAI leverage the natural language processing capabilities of Large Language Models to generate machine code from natural language descriptions. In the cybersecurity field, red teams can also take advantage of generative models to build malicious code generators, providing more automation to Pentest audits. However, the application of Large Language Models in malicious code generation remains challenging due to the lack of data to train and evaluate offensive code generators. In this work, we propose RedShell, a tool that allows ethical hackers to generate malicious PowerShell code. We also introduce a ground truth dataset, combining publicly available code samples to fine-tune models in malicious PowerShell generation. Our experiments demonstrate the strong capabilities of RedShell in generating syntactically valid PowerShell, with fewer than 10\% of the generated samples resulting in parse errors. Furthermore, our specialized model was able to produce samples that were semantically consistent with reference snippets, achieving a competitive performance on standard output similarity metrics such as Edit Distance and METEOR, with their mean similarity scores exceeding 50\% and 40\%, respectively. This work sheds light on the state-of-the-art research in the field of Generative AI applied to Pentesting, and also serves as a steppingstone for future advancements, highlighting the potential benefits these models hold within such controlled environments. 

\end{abstract}

\section{Introduction}

In recent years, Large Language Models (LLMs) sparked a revolution in natural language processing through advanced transformers~\cite{Vaswani2017}. Generative models such as ChatGPT~\cite{chatgpt} have demonstrated strong capabilities while performing generic tasks~\cite{Wolf2020}. In addition, fine-tuning techniques have been employed to adapt LLMs to address more specific challenges, with Generative Artificial Intelligence (AI) becoming pervasive in the workflow of many software engineering tasks~\cite{Fan2023}. 

In the offensive cybersecurity field, ethical hackers can also take advantage of LLMs to develop malicious code generators, providing more automation to Pentest audits. Penetration Testing, often referred to as Pentesting, is a crucial activity for red teams, which are responsible for testing the cybersecurity effectiveness of a system through simulated cyber-attacks~\cite{Vats2020}. By detecting potential security lapses in a target system, Pentesters are able to prevent attacks that could cause harm to that system and its users. 

Ethical hackers typically use malicious code to exploit the previously identified vulnerabilities, assessing what real attackers could gain after a successful intrusion. However, regardless of the exploitation technique being employed, writing offensive code requires time and technical knowledge from the Pentesters. While the desired automation could be provided by LLMs, this strategy also remains challenging due to the lack of data to train offensive code generators.

In this work, we propose RedShell, an AI-based malicious PowerShell generator designed to provide more automation to Pentesting activities targeting Microsoft Windows. We also introduce a malicious PowerShell ground truth dataset, combining code samples from various cybersecurity frameworks to train and evaluate models in offensive PowerShell generation.

The evaluation focuses on gauging the quality of the generated snippets by examining their syntactic and semantic correctness, as well as performing a comparative analysis of the generation capabilities of different LLMs. Experiments show that our tool was able to generate syntactically valid PowerShell code, with fewer than 10\% of the generated samples resulting in parse errors. Additionally, RedShell produced samples closely aligned with reference snippets, achieving high scores in common distance metrics such as Edit Distance and METEOR, with their mean similarity scores exceeding 50\% and 40\%, respectively.

The remainder of this article is organized as follows: Section~\ref{sec:ground_truth} introduces an extended version of a malicious PowerShell dataset from the literature; Section~\ref{sec:redshell} details the design of RedShell, our specialized LLM fine-tuned on this dataset; Section~\ref{sec:tests} presents experiments validating RedShell’s generation capabilities; Section~\ref{sec:related_work} reviews related work, and finally, Section~\ref{sec:conclusions} concludes the article with a discussion of RedShell’s limitations and directions for future work.

\section{Ground Truth Dataset}
\label{sec:ground_truth}

Offensive code datasets are a crucial component for red teams to develop malicious code generators. However, there is a significant gap in the public availability of offensive code samples. Without collecting the required amount of snippets, the training and evaluation of malicious code generators is highly compromised. 

To develop RedShell, our approach was to build a ground truth dataset combining offensive PowerShell snippets and their corresponding descriptions in the English language. We built the ground truth dataset in two distinct phases, presenting the reference ground truth and the extended ground truth datasets.

The reference ground truth dataset corresponds to the offensive PowerShell collection provided by~\citeauthor{Liguori2024} in \cite{Liguori2024}. The dataset is composed of 1,127 samples of offensive PowerShell, capturing different real-world scenarios where Pentesters may take advantage of PowerShell to target Microsoft Windows.

The snippets comprising the reference dataset were collected from community-driven cybersecurity wikis and blogs about ethical hacking such as Red Team Recipe~\cite{redteamrecipe}. Additionally, snippets were also extracted from various cybersecurity frameworks, including Atomic Red Team~\cite{AtomicRedTeam}. The reference ground truth dataset covers 13 out of 14 offensive tactics described by MITRE ATT\&CK~\cite{mitre_attack}, only leaving uncovered the \textit{Resource Development} tactic.

We built an extended version of the reference ground truth dataset by collecting new malicious samples, thus improving the overall quality and size of the dataset. We introduced 1,135 additional code samples of offensive PowerShell, effectively doubling the size of the original dataset. The collected samples were extracted from offensive PowerShell modules such as Mimikatz~\cite{Mimikatz} and Nishang~\cite{nishang}, and from tryhackme~\cite{TryHackMe} walkthroughs on capture the flag challenges.

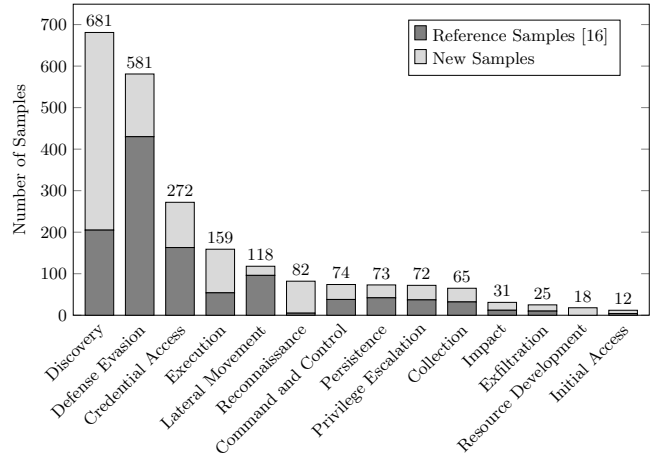
\begin{figure}[htbp]
    \centering
    \begin{adjustbox}{width=0.7\textwidth}
    \begin{tikzpicture}
        \begin{axis}[ 
            ybar stacked,  
            bar width=15pt,
            legend style={at={(0.95,0.95)}, anchor=north east, font=\small, /tikz/every node/.style={anchor=west}},
            symbolic x coords={Discovery, Defense Evasion, Credential Access, Execution, Lateral Movement, Reconnaissance, Command and Control, Persistence, Privilege Escalation, Collection, Impact, Exfiltration, Resource Development, Initial Access},
            xtick=data,
            x tick label style={rotate=45, anchor=east, font=\small, yshift=-5pt},
            ymin=0,
            ytick={0, 100, 200, 300, 400, 500, 600, 700},
            ylabel={Number of Samples},
            width=\textwidth,
            height=0.6\textwidth,
            enlarge x limits=0.05,
            xtick style={draw=none},
        ]

            \addplot[fill=gray] coordinates {
                (Discovery, 205)
                (Defense Evasion, 430)
                (Credential Access, 163)
                (Execution, 54)
                (Lateral Movement, 96)
                (Reconnaissance, 5)
                (Command and Control, 38)
                (Persistence, 42)
                (Privilege Escalation, 37)
                (Collection, 32)
                (Impact, 12)
                (Exfiltration, 10)
                (Resource Development, 0)
                (Initial Access, 4)
            };

            \addplot[fill=gray!30] coordinates {
                (Discovery, 476)
                (Defense Evasion, 151)
                (Credential Access, 109)
                (Execution, 105)
                (Lateral Movement, 22)
                (Reconnaissance, 77)
                (Command and Control, 36)
                (Persistence, 31)
                (Privilege Escalation, 35)
                (Collection, 33)
                (Impact, 19)
                (Exfiltration, 15)
                (Resource Development, 18)
                (Initial Access, 8)
            };

            \node at (axis cs:Discovery,681) [anchor=south, font=\footnotesize] {681};
            \node at (axis cs:Defense Evasion,581) [anchor=south, font=\footnotesize] {581};
            \node at (axis cs:Credential Access,272) [anchor=south, font=\footnotesize] {272};
            \node at (axis cs:Execution,159) [anchor=south, font=\footnotesize] {159};
            \node at (axis cs:Lateral Movement,118) [anchor=south, font=\footnotesize] {118};
            \node at (axis cs:Reconnaissance,82) [anchor=south, font=\footnotesize] {82};
            \node at (axis cs:Command and Control,74) [anchor=south, font=\footnotesize] {74};
            \node at (axis cs:Persistence,73) [anchor=south, font=\footnotesize] {73};
            \node at (axis cs:Privilege Escalation,72) [anchor=south, font=\footnotesize] {72};
            \node at (axis cs:Collection,65) [anchor=south, font=\footnotesize] {65};
            \node at (axis cs:Impact,31) [anchor=south, font=\footnotesize] {31};
            \node at (axis cs:Exfiltration,25) [anchor=south, font=\footnotesize] {25};
            \node at (axis cs:Resource Development,18) [anchor=south, font=\footnotesize] {18};
            \node at (axis cs:Initial Access,12) [anchor=south, font=\footnotesize] {12};

            \legend{Reference Samples~\cite{Liguori2024}, New Samples}
        \end{axis}
    \end{tikzpicture}
    \end{adjustbox}
    \caption{Ground truth dataset coverage of the MITRE ATT\&CK tactics.}
    \label{fig:mitre-mapping}
\end{figure}

While we manually classified the offensive tactic of each new collected sample according to MITRE ATT\&CK documentation, for the reference dataset we relied on the classification provided by~\citeauthor{Liguori2024} in \cite{Liguori2024}. Figure~\ref{fig:mitre-mapping} presents the number of collected PowerShell samples for each tactic, highlighting how both reference and extended datasets contribute to the final ground truth composition. 

The new collected samples improved significantly the dataset coverage of the offensive tactics typically employed by Pentesters, including \textit{Discovery}, \textit{Defense Evasion}, and \textit{Credential Access}. The high availability of \textit{Discovery} samples highlights the capabilities offered by PowerShell to perform various discovery activities on Microsoft Windows devices such as permission groups discovery. 

The ground truth dataset also provides a wide representation of \textit{Defense Evasion} strategies, employed by ethical hackers to avoid detection and conceal traces of their malicious activities. Evasion techniques include disabling or modifying security software such as the Windows Defender and the Antimalware Scan Interface (AMSI), to allow the stealthy execution of malicious programs.

Additionally, security professionals take advantage of \textit{Credential Access} strategies to dump credentials and steal account names and passwords. Legitimate credentials can then be used to access sensitive information and services while making malicious activities harder to detect.

\section{RedShell Design}
\label{sec:redshell}

To develop RedShell, we followed a methodology where the malicious PowerShell dataset described in Section~\ref{sec:ground_truth} was employed as a knowledge base to fine-tune and compare three different LLMs.

\subsection{Models}

We selected LLMs based on the following criteria: \textit{i}) strong performance in coding tasks such as code generation, summarization, and reasoning; \textit{ii}) representation of the state-of-the-art model families and architectures; \textit{iii}) public availability of model weights to allow local fine-tuning; and \textit{iv}) relatively small model sizes to support efficient experimentation. The chosen models are:

\begin{description}
  \item[Qwen2.5-7B~\cite{qwen25}] Belongs to the latest series of Qwen LLMs, which offer significantly more knowledge than the previously released versions while providing improved capabilities in coding and mathematics.
  \vspace{2mm}
  \item[Qwen2.5-Coder-7B-Instruct~\cite{qwenCoder}] Belongs to the latest series of code-specific Qwen LLMs, offering strong capabilities in code generation, reasoning and fixing.
  \vspace{2mm}
  \item[Llama3.1-8B~\cite{llama}] Released on July of 2024 as part of the multilingual collection of LLMs from Meta, outperforming many of the available open-source and closed models on common industry coding benchmarks. 
\end{description}

The selected models were downloaded from HuggingFace~\cite{huggingface} and fine-tuned locally, providing properties to our solution that overcome the limitations presented by closed models such as ChatGPT from OpenAI. When compared to proprietary LLMs, RedShell offers the following properties:

\begin{description}
  \item[Privacy.]  By manipulating our specialized models locally, we avoid to share sensitive data with private companies whose proprietary LLMs are only accessible through an external API.
  \vspace{3mm}
  \item[Specialization.] Our specialized models were specifically trained to assist Pentesters in malicious PowerShell generation. The same training process could not be applied to closed LLMs since their weights are not publicly accessible.
  \vspace{3mm}
  \item[Ethical Boundaries.] Closed LLMs are usually protected by ethical boundaries that restrict their generation capabilities. Although these protections can be often bypassed through prompt-engineering techniques, that manipulation requires time and effort from the Pentesters. In contrast, the generation capabilities of our specialized LLMs are not limited by ethical boundaries. While unrestricted LLMs may raise ethical concerns, attackers will inevitably exploit them, making it essential for security professionals to do the same to effectively prevent real threats. Additionally, to ensure the responsible use of RedShell, we excluded from its training data all the \textit{Impact} samples that could potentially compromise the integrity of target systems, thus aligning our tool with the non-destructive nature of Pentesting.
\end{description}

\subsection{Fine-Tuning}

The training of the selected LLMs was conducted through Unsloth~\cite{unsloth}, a fine-tuning framework that manually patches complex mathematical steps and optimizes GPU kernels and VRAM (Video Random Access Memory) allocation to make training faster without any hardware changes. Unsloth also supports partial fine-tuning processes through the Low Rank Adaption Method (LoRA)~\cite{lora}, which allows the training to adjust only a small number of weights from the selected models, reducing the computational costs of the fine-tune. 

We adopted standard values for both LoRA and training parameters, following the recommendations provided by the Unsloth documentation. Additionally, to determine the most effective training configuration, we conducted a set of experiments evaluating the performance of models fine-tuned under different settings. We describe the experiments in more detail in Subsection~\ref{sec:eval}. In particular, we parametrized LoRA with both \textit{rank} and \textit{alpha} equal to 64, and a 0 \textit{dropout}. For the fine-tuning process, we employed a \textit{batch size} equal to 8 and a \textit{learning rate} equal to $2 \times e^{-4}$. We also provided the LLMs with a context that included a detailed description of their expected behavior. The context was defined as \textit{``Act as a malicious PowerShell generator. Generate commands in a single line, separated by semicolons and provide no further explanations''}.

The ground truth dataset was randomly split in two partitions, the training and the test datasets. Since the fine-tuning was conducted with a small and manually curated dataset, we adopted a 90/10 train-test split to maximize the amount of data available for training, optimizing the intended knowledge transfer. The test dataset, while comprising a significantly smaller part of ground truth, still provides a sufficient amount of unseen data to assess the generalization capabilities of the fine-tuned models and detect potential overfitting scenarios. Since the reference dataset provided only aggregate tactic counts rather than a tactical classification per snippet, the train-test split does not guarantee balanced representation across MITRE ATT\&CK tactics. However, this does not represent a threat to the validity of RedShell since the most common Pentesting tactics such as \textit{Discovery}, \textit{Defense Evasion} and \textit{Credential Access} dominate the dataset, meaning that random partitioning still ensures a high likelihood of their presence on both partitions.

The fine-tuning processes were executed on a local Linux machine, employing a single NVIDIA GeForce RTX 4090 GPU and 23.643 GB of VRAM. By taking advantage of the optimizations provided by Unsloth and LoRA, our strategy minimizes the time, energy and computational resources required to fine-tune the selected LLMs. Table~\ref{tab:train} presents the training times of our specialized models (both in number of epochs and minutes), and the peak reserved VRAM for each model fine-tuned with the reference ground truth dataset. We can perceive that the observed VRAM peaks are significantly far below our max VRAM limit of 23.643 GB. Additionally, the number of epochs (i.e. complete model iterations through the training data) was defined based on experimentation (Subsection~\ref{sec:eval}), reflecting the different learning abilities of each LLM.
\vspace{-4mm}
\begin{table}[htbp!]
    \caption{Training report of LLMs fine-tuned with the reference dataset.}
    \label{tab:train}
    \centering
    \renewcommand{\arraystretch}{1.3}
    \renewcommand{\theadfont}{\bfseries}
    \hyphenpenalty=10000
     \begin{tabularx}{\textwidth}{
        l
        >{\centering\arraybackslash}X
        >{\centering\arraybackslash}X
        >{\centering\arraybackslash}X
    }
    \toprule
    \thead{Model} & \thead{Epochs} & \thead{Total Training \\Time (min)} & \thead{Peak Reserved\\VRAM (GB)}\\
    \midrule
            Llama-3.1 (8B) & 18 & 28 & 16.725 \\
            Qwen2.5-Coder (7B) & 20 & 30 & 16.865 \\
            Qwen2.5 (7B) & 28 & 47 &  17.600 \\
    \bottomrule
    \end{tabularx}
\end{table}

\subsection{Evaluation}
\label{sec:eval}

An experimental evaluation was conducted to assess the effectiveness of employing the fine-tuned LLMs as malicious PowerShell generators. We inferred our specialized models using the previously unseen code descriptions from the test dataset. The evaluation aimed to validate the quality of the generated code blocks by examining their syntactic and semantic correctness.  

The syntactic correctness of the PowerShell snippets was evaluated based on the number and severity of the syntactic flaws identified by PSScriptAnalyzer~\cite{PSScriptAnalyzer}, a static PowerShell code checker provided by Microsoft. The reported occurrences were the following:
\vspace{-1.5mm}
\begin{description}
  \item[Parse Errors.] High-severity errors that occur during the parsing of the PowerShell code, preventing the execution of the generated samples.
  \vspace{3mm}
  \item[Warnings.] Flaws that may alert for the presence of bad coding practices or unexpected PowerShell patterns.
  \vspace{3mm}
  \item[Errors.] High-severity flaws that alert for the violation of semantic and security rules from PowerShell.
\end{description}

The presence of parse errors in the generated samples was a crucial metric to identify the snippets that could not be executed. In contrary, PowerShell warnings and errors typically do not prevent the code from executing. However, these occurrences allowed us to evaluate the quality of the generated samples in terms of the adherence to the best PowerShell practices. 

The results from the syntactic evaluation were then used to compute the parse error, warning, and error percentages in the samples generated by our specialized models. Since a single sample could potentially contain multiple parse errors, warnings and errors, simultaneously, our approach was to classify a sample as having a parse error if the code for that sample registered one or more parse errors, regardless of the additional presence of warnings or errors. Samples that did not register parse errors were then classified as containing warnings or errors if their code included at least one warning or one error, respectively. Samples (without parse errors) containing both warnings and errors were classified under both categories.

Additionally, we defined the semantic correctness of the PowerShell code as a mean distance that statistically measures the similarity between the generated snippets and the corresponding expected snippets in the test dataset. To measure the semantic correctness of the PowerShell samples, we computed the standard output similarity metrics, described by~\citeauthor{Liguori2023} in~\cite{Liguori2023}:

\begin{description}
  \item[ROUGE-L] Measures the similarity between the reference and generated code samples based on the longest common subsequence metric, producing a score that ranges between 0 (perfect mismatch) and 1 (perfect matching). We employed the \textit{rouge}~\cite{rouge} Python package to compute ROUGE-L.
  \vspace{3mm}
  \item[METEOR] Measures the alignment between reference and generated samples by mapping unigrams, producing a score that ranges between 0 (perfect mismatch) and 1 (perfect matching). We computed METEOR by leveraging the \textit{evaluate}~\cite{evaluate} Python package from HuggingFace.
  \vspace{3mm}
  \item[BLEU] Measures the $n$-gram intersection between the reference and generated snippets using a score that ranges between 0 (perfect mismatch) and 1 (perfect matching), penalizing the score of the generated samples that are longer than their corresponding references. We computed BLEU for $n$-grams with $n = 4$, taking advantage of the BLEU implementation provided by Microsoft in CodeXGLUE~\cite{bleu}, a benchmark dataset on code intelligence.
  \vspace{3mm}
  \item[Edit Distance] Measures the output distance by computing the minimum number of operations on single characters required to make each generated snippet equal to the reference sample. The ED score ranges between 0 (perfect matching) and a positive integer representing the number of character operations required to achieve a perfect matching. We computed ED through the \textit{pylcs}~\cite{pylcs} Python package, normalizing the produced score between 0 (perfect mismatch) and 1 (perfect matching).
  \vspace{3mm}
  \item [Exact-Match] Measures the mean percentage of generated samples that perfectly match their corresponding reference samples in test dataset.
\end{description}

\section{Syntactic and Semantic Assessment}
\label{sec:tests}

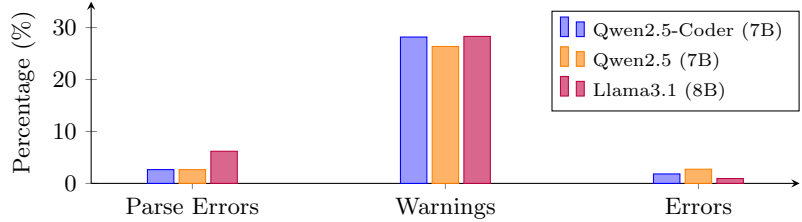
\begin{figure}[ht]
\centering
\begin{minipage}{\textwidth}
\begin{tikzpicture}
\begin{axis}[
    ybar,
    bar width=10pt,
    enlargelimits=0.15,
    ymin=0, ymax=35,
    ytick={0,10,20,30},
    ylabel={Percentage (\%)},
    symbolic x coords= {Parse Errors, Warnings, Errors},
    xtick=data,
    xticklabel style={rotate=0},
    axis x line=bottom,
    axis y line=left,
    legend style={at={(0.5,-0.2)}, anchor=north, legend columns=3},
    width=11cm,
    height=4cm,
    grid=none,
    enlarge x limits=0.2,
    legend style={
        at={(1,0.95)},
        anchor=north east,
        legend columns=1,    
        font=\scriptsize,
        /tikz/every node/.style={anchor=west}
    },
]

\addplot+[style={blue, fill=blue!40}] coordinates {
    (Parse Errors, 2.65)
    (Warnings, 28.18)
    (Errors, 1.82)
};

\addplot+[style={orange, fill=orange!60}] coordinates {
    (Parse Errors, 2.65)
    (Warnings, 26.36)
    (Errors, 2.73)
};

\addplot+[style={purple, fill=purple!60}] coordinates {
    (Parse Errors, 6.19)
    (Warnings, 28.30)
    (Errors, 0.94)
};

\legend{Qwen2.5-Coder (7B), Qwen2.5 (7B), Llama3.1 (8B)}
\end{axis}
\end{tikzpicture}  
\vspace{-2mm}
\caption{Syntactic evaluation of models fine-tuned with the reference dataset.}
\label{fig:syntax}
\end{minipage}
\end{figure}

The results of the syntactic evaluation of the models fine-tuned with the reference ground truth dataset can be observed in Figure~\ref{fig:syntax}. According to PSScriptAnalyzer, our specialized models were able to generate valid malicious PowerShell code. In fact, the generated samples registered low parse error and error percentages, highlighting the strong capabilities of the fine-tuned LLMs in generating syntactically correct PowerShell code.

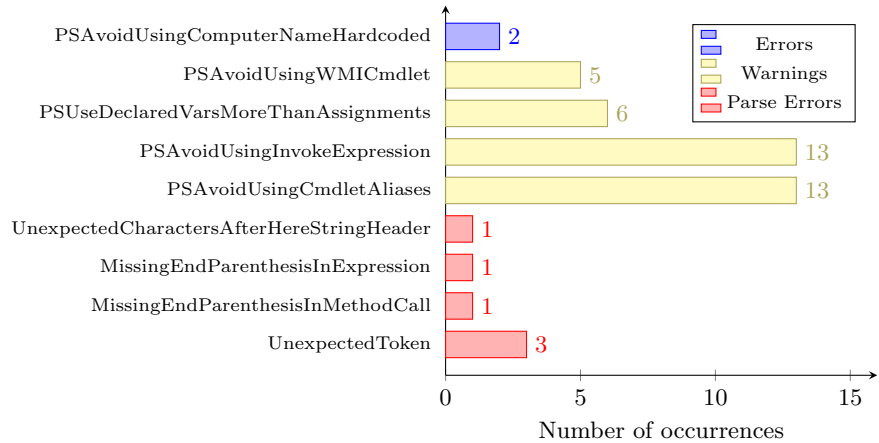
\begin{figure}[htbp]
\centering
\begin{tikzpicture}
\begin{axis}[
    xbar,
    bar width=10pt,
    xlabel={Number of occurrences},
    xmin=0, xmax=16,
    ymin=0, ymax=8,
    xtick={0,5,10,15},
    ytick={0,...,8},
    yticklabels={
        UnexpectedToken,
        MissingEndParenthesisInMethodCall,
        MissingEndParenthesisInExpression,
        UnexpectedCharactersAfterHereStringHeader,
        PSAvoidUsingCmdletAliases,
        PSAvoidUsingInvokeExpression,
        PSUseDeclaredVarsMoreThanAssignments,
        PSAvoidUsingWMICmdlet,
        PSAvoidUsingComputerNameHardcoded
    },
    yticklabel style={font=\scriptsize, align=right, inner ysep=0pt},
    nodes near coords,
    nodes near coords align={horizontal},
    every node near coord/.append style={font=\footnotesize},
    width=0.6\textwidth,
    height=6.5cm,
    axis y line=left,
    axis x line=bottom,
    legend style={at={(0.95,0.95)}, anchor=north east, font=\scriptsize},
    xbar legend,
    ytick style={draw=none},
    bar shift=0pt,
    enlarge y limits=0.1
]

\newcommand{\plotpos}[1]{
    \ifcase#1\relax
    \or 0 
    \or 1 
    \or 2 
    \or 3 
    \or 4 
    \or 5 
    \or 6 
    \or 7 
    \or 8 
    \fi
}

\addplot+[blue, fill=blue!30] coordinates {
    (2,\plotpos{9})
};
\addlegendentry{Errors}

\addplot+[yellow!60!black, fill=yellow!30] coordinates {
    (13,\plotpos{5})
    (13,\plotpos{6})
    (6,\plotpos{7})
    (5,\plotpos{8})
};
\addlegendentry{Warnings}

\addplot+[red, fill=red!30] coordinates {
    (3,\plotpos{1})
    (1,\plotpos{2})
    (1,\plotpos{3})
    (1,\plotpos{4})
};
\addlegendentry{Parse Errors}

\end{axis}
\end{tikzpicture}
\caption{Syntax report of Qwen2.5-Coder fine-tuned with the reference dataset.}
\label{fig:qwen-coder-syntax}
\end{figure}

From the three fine-tuned LLMs, Qwen2.5 and Qwen2.5-Coder were the models that produced samples with lower parse error percentages. Additionally, all the three specialized models produced snippets with significantly high warning percentages. However, since the LLMs were specifically trained to produce malicious code, it was expected that PSScriptAnalyzer would identify warnings in the generated samples regarding safety violations of PowerShell. For instance, Figure~\ref{fig:qwen-coder-syntax} presents all the syntax flaws identified by PSScriptAnalyzer in the samples generated by Qwen2.5-Coder fine-tuned with the reference ground truth dataset. One of the most registered warnings was \textit{``PSAvoidUsingInvokeExpression''}, which alerts for the usage of a PowerShell command that enables the execution of potentially unsafe code via string representation.

The \textit{``UnexpectedToken''} parse error was the most frequent occurrence to prevent the execution of the generated samples. Furthermore, \textit{``PSAvoidUsingComputerNameHardcoded''} was the most registered PowerShell error. According to PSScriptAnalyzer, hardcoding the value of the \textit{ComputerName} argument violates a security rule from PowerShell since it will potentially expose sensitive information regarding the target host.

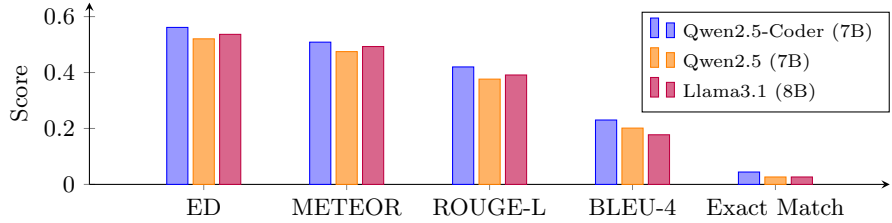
\begin{figure}[ht]
\centering
\begin{tikzpicture}
\begin{axis}[
    ybar,
    bar width=8pt,
    enlargelimits=0.15,
    ymin=0, ymax=0.65,
    ytick={0,0.2,0.4,0.6},
    ylabel={Score},
    symbolic x coords={ED, METEOR, ROUGE-L, BLEU-4, Exact Match},
    xtick=data,
    axis x line=bottom,
    axis y line=left,
    enlarge x limits=0.2,
    legend style={
        at={(1,0.95)},
        anchor=north east,
        legend columns=1,    
        font=\scriptsize,
        /tikz/every node/.style={anchor=west}
    },
    width=14cm,
    height=4cm,
    grid=none,
    enlarge x limits=0.2,
    x=1.9cm,
]

\addplot+[style={\qwencoderB,fill=\qwencoderF}] coordinates {
    (BLEU-4, 0.2302)
    (ROUGE-L, 0.4200)
    (Exact Match, 0.0442)
    (ED, 0.5613)
    (METEOR, 0.5086)
};

\addplot+[style={\qwenB,fill=\qwenF}] coordinates {
    (BLEU-4, 0.2013)
    (ROUGE-L, 0.3764)
    (Exact Match, 0.0265)
    (ED, 0.5205)
    (METEOR, 0.4748)
};

\addplot+[style={\llamaB,fill=\llamaF}] coordinates {
    (BLEU-4, 0.1774)
    (ROUGE-L, 0.3912)
    (Exact Match, 0.0265)
    (ED, 0.5365)
    (METEOR, 0.4930)
};

\legend{Qwen2.5-Coder (7B), Qwen2.5 (7B), Llama3.1 (8B)}
\end{axis}
\end{tikzpicture}
\caption{Semantic evaluation of models fine-tuned with the reference dataset.}
\label{fig:semantic}
\end{figure}

Figure~\ref{fig:semantic} presents the results of the semantic evaluation performed for our specialized models. The fine-tuned LLMs achieved significantly high scores across different output similarity metrics, highlighting their strong capabilities to generate malicious PowerShell closely aligned with the expected code references. In particular, Qwen2.5-Coder outperformed Qwen2.5 and Llama3.1 in all output similarity metrics.

\begin{figure}[ht]
\centering
\begin{tikzpicture}
\begin{axis}[
    ybar,
    bar width=8pt,
    enlargelimits=0.15,
    ymin=0, ymax=0.65,
    ytick={0,0.2,0.4,0.6},
    ylabel={Score},
    symbolic x coords={ED, METEOR, ROUGE-L, BLEU-4, Exact Match},
    xtick=data,
    axis x line=bottom,
    axis y line=left,
    enlarge x limits=0.2,
    legend style={
        at={(1,1)},
        anchor=north east,
        legend columns=1,    
        font=\scriptsize,
        /tikz/every node/.style={anchor=west}
    },
    width=14cm,
    height=4cm,
    grid=none,
    enlarge x limits=0.2,
    x=1.9cm,
]

\addplot+[style={\qwencoderB,fill=\qwencoderF}] coordinates {
    (BLEU-4, 0.2849)
    (ROUGE-L, 0.4200)
    (Exact Match, 0.0442)
    (ED, 0.5613)
    (METEOR, 0.5086)
};

\addplot+[style={\codegptB,fill=\codegptF}] coordinates {
    (BLEU-4, 0.2171)
    (ROUGE-L, 0.3863)
    (Exact Match, 0.0354)
    (ED, 0.5017)
    (METEOR, 0.4534)
};

\addplot+[style={\codegenB,fill=\codegenF}] coordinates {
    (BLEU-4, 0.1853)
    (ROUGE-L, 0.3545)
    (Exact Match, 0.0177)
    (ED, 0.4867)
    (METEOR, 0.4414)
};

\addplot+[style={\codetfiveB,fill=\codetfiveF}] coordinates {
    (BLEU-4, 0.1850)
    (ROUGE-L, 0.3886)
    (Exact Match, 0.0177)
    (ED, 0.5023)
    (METEOR, 0.4787)
};

\legend{Qwen2.5-Coder, CodeGPT~\cite{Liguori2024}, CodeGen~\cite{Liguori2024}, CodeT5+~\cite{Liguori2024}}
\end{axis}
\end{tikzpicture}
\caption{Semantic evaluation of Qwen2.5-Coder and reference models.}
\label{fig:semantic-ref}
\end{figure}
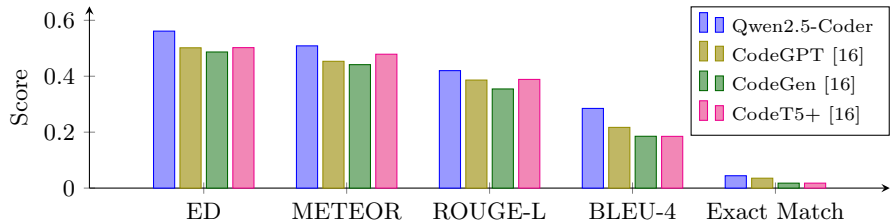

To further validate the generation capabilities of our specialized models, we conducted a detailed analysis of the state-of-the-art solutions for automatic malicious PowerShell generation through LLMs. We used as reference the specialized versions of CodeGPT, CodeGen and CodeT5+ proposed in~\cite{Liguori2024}. The reference models were fine-tuned by~\citeauthor{Liguori2024} using the reference ground truth dataset. Additionally, CodeT5+ and CodeGen were also pre-trained with a dataset composed of 89,814 generic PowerShell samples extracted from GitHub. Figure~\ref{fig:semantic-ref} presents the comparative analysis of the semantic evaluation results achieved by the reference models (reported in~\cite{Liguori2024}) and our specialized version of Qwen2.5-Coder, also fine-tuned with the reference ground truth dataset.

Qwen2.5-Coder outperformed the reference models across all output similarity metrics. Notably, our specialized LLM was only partially fine-tuned through LoRA and Unsloth, using a single GPU under low VRAM consumption conditions, while the reference models were targeted by a complete fine-tune, with CodeT5+ and CodeGen also benefiting from a specialized pre-train in generic PowerShell.

\begin{figure}[ht]
\centering
\begin{tikzpicture}
\begin{axis}[
    ybar,
    bar width=8pt,
    enlargelimits=0.15,
    ymin=0, ymax=0.65,
    ytick={0,0.2,0.4,0.6},
    ylabel={Score},
    symbolic x coords={ED, METEOR, ROUGE-L, BLEU-4, Exact Match},
    xtick=data,
    axis x line=bottom,
    axis y line=left,
    enlarge x limits=0.2,
    legend style={
        at={(1,0.95)},
        anchor=north east,
        legend columns=1,    
        font=\scriptsize,
        /tikz/every node/.style={anchor=west}
    },
    width=14cm,
    height=4cm,
    grid=none,
    x=1.9cm,
]

\addplot+[style={\qwencoderB,fill=\qwencoderF}] coordinates {
    (BLEU-4, 0.2302)
    (ROUGE-L, 0.4200)
    (Exact Match, 0.0442)
    (ED, 0.5613)
    (METEOR, 0.5086)
};

\addplot+[style={\chatgptB,fill=\chatgptF}] coordinates {
    (BLEU-4, 0.0753)
    (ROUGE-L, 0.2313)
    (Exact Match, 0.0088)
    (ED, 0.3384)
    (METEOR, 0.2217)
};

\addplot+[style={\deepseekB,fill=\deepseekF}] coordinates {
    (BLEU-4, 0.0683)
    (ROUGE-L, 0.2276)
    (Exact Match, 0.0088)
    (ED, 0.3612)
    (METEOR, 0.2632)
};

\legend{Qwen2.5-Coder, ChatGPT 3.5, DeepSeekChat-V3}
\end{axis}
\end{tikzpicture}
\caption{Semantic evaluation of Qwen2.5-Coder and closed models.}
\label{fig:popular}
\end{figure}
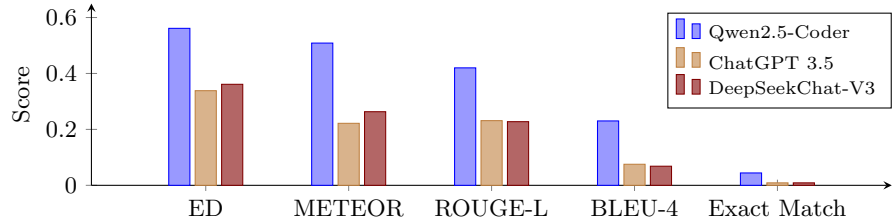

We also compared the semantic performance of Qwen2.5-Coder with some of the most popular proprietary LLMs such as ChatGPT 3.5~\cite{chatgpt} and DeepSeekChat-V3~\cite{deepseek}, as illustrated in Figure~\ref{fig:popular}. While we inferred DeepSeekChat through its public API and evaluated the generated snippets, for the ChatGPT assessment we relied on the scores computed by~\citeauthor{Liguori2024} in~\cite{Liguori2024} using the same code descriptions. Achieved results demonstrate that Qwen2.5-Coder is a strong alternative for malicious PowerShell generation when compared to the current state-of-the-art solutions.

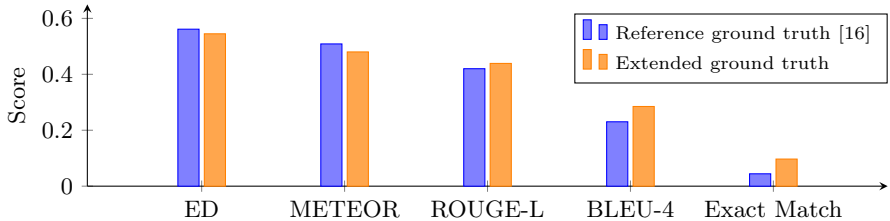
\begin{figure}[ht]
\centering
\begin{tikzpicture}
\begin{axis}[
    ybar,
    bar width=8pt,
    enlargelimits=0.15,
    ymin=0, ymax=0.65,
    ytick={0,0.2,0.4,0.6},
    ylabel={Score},
    symbolic x coords={ED, METEOR, ROUGE-L, BLEU-4, Exact Match},
    xtick=data,
    axis x line=bottom,
    axis y line=left,
    enlarge x limits=0.2,
    legend style={
        at={(1,0.95)},
        anchor=north east,
        legend columns=1,    
        font=\scriptsize,
        /tikz/every node/.style={anchor=west}
    },
    width=14cm,
    height=4cm,
    grid=none,
    x=1.9cm,
]

\addplot+[style={blue,fill=blue!50}] coordinates {
    (BLEU-4, 0.2302)
    (ROUGE-L, 0.4200)
    (Exact Match, 0.0442)
    (ED, 0.5613)
    (METEOR, 0.5086)
};

\addplot+[style={orange,fill=orange!70}] coordinates {
    (BLEU-4, 0.2849)
    (ROUGE-L, 0.4389)
    (Exact Match, 0.0969)
    (ED, 0.5448)
    (METEOR, 0.4800)
};

\legend{Reference ground truth~\cite{Liguori2024}, Extended ground truth}
\end{axis}
\end{tikzpicture}
\vspace{-2mm}
\caption{Semantic evaluation of Qwen2.5-Coder fine-tuned with different datasets.}
\label{fig:combined}
\end{figure}

Up to this point, conducted experiments focused on evaluating the selected models fine-tuned only with the reference ground truth dataset from~\cite{Liguori2024}. To assess the training benefits of employing our extended dataset (previously introduced in Section \ref{sec:ground_truth}), we compared the semantic evaluation performance of two different versions of Qwen2.5-Coder, one fine-tuned only with the original reference dataset, and the other fine-tuned with the extended ground truth dataset. As illustrated in Figure~\ref{fig:combined}, the ED, METEOR and ROUGE-L scores exhibited negligible variation between the two versions of Qwen2.5-Coder. This happens since these metrics focus on overall structure and semantic similarity, areas where the model fine-tuned with the reference dataset already performed well. In contrast, the model fine-tuned with our extended dataset demonstrated more substantial improvements in BLEU-4 and Exact Match, stricter metrics that emphasize longer n-gram overlaps and exact matches. This highlights the positive impact of using a larger and more diverse training dataset, which primarily enhanced output precision rather than general structure or meaning.

Based on the previous results, we defined the specialized version of Qwen2.5-Coder as the best candidate to be incorporated in RedShell. Table~\ref{tab:snippets} compares some illustrative examples of snippets generated by our tool and the corresponding ground truth references, visually emphasizing their strong similarity. Code differences are highlighted in red, where slashed text denotes unexpected elements.

\begin{table}[htp]
    \caption{Illustrative examples of snippets generated by RedShell.}
    \label{tab:snippets}
    \renewcommand{\arraystretch}{1.3}
    \hyphenpenalty=10000
    \centering
    \begin{tabularx}{\textwidth}{>{\RaggedRight}X >{\RaggedRight}X >{\RaggedRight}X}
        \toprule
        \textbf{Label} & \textbf{Reference Snippet} & \textbf{Generated Snippet} \\
        \midrule
        List the members of Admins group. & 
        \texttt{Get-ADGroupMember -Identity "Admins"} &  
        \texttt{Get-ADGroupMember -Identity "Admins"} \\ \midrule[0.1pt]
        Use Mimikatz to dump plain text passwords from memory. & 
        \texttt{Invoke-Mimikatz -Command '"sekurlsa:: logonpasswords"'} & 
        \texttt{Invoke-Mimikatz -Command '"sekurlsa:: logonpasswords"' \textcolor{red}{\textbar\ \cancel{Out-File -FilePath} C:\textbackslash temp\textbackslash creds.txt}} \\ \midrule[0.1pt]
        Execute a malicious command encoded as a Base64 string. & \texttt{powershell.exe -c "iex ([System.Text.Encoding] ::Unicode.GetString( [System.Convert]:: FromBase64String( 'Cmd')))"} & 
        \texttt{\textcolor{red}{powershell.exe -c"} iex \textcolor{red}{\cancel{-Command}} ([System.Text.Encoding] ::\textcolor{red}{\cancel{UTF8}}.GetString( [System.Convert]::  FromBase64String ('\textcolor{red}{\cancel{base64Encoded}}Cmd')))\textcolor{red}{"}} \\
        \bottomrule
    \end{tabularx}
    \vspace{3mm}
\end{table}

\section{Related Work}
\label{sec:related_work}

LLMs have registered a high popularity in recent years, with the AI-community actively searching for new approaches to create small, and yet powerful, generative models. In addition, engineers are taking advantage of these novel technologies to provide more automation to complex and time-consuming tasks.

In the offensive cybersecurity field, various AI-based solutions have been proposed to assist ethical hackers in Pentesting scenarios. Tools such as PentestGPT~\cite{Deng2023} and PENTEST-AI~\cite{Bianou2024} aim to provide full automation to Pentest audits through complex LLM agents, providing mechanisms to detect, explore, and report different vulnerabilities with minimal human intervention. In contrast, RedShell provides a solution specifically designed to assist Pentesters in malicious PowerShell generation. 

Some literature also focused on the application of LLMs in offensive code generation.~\citeauthor{Liguori2021} in~\cite{Liguori2021} proposed EVIL, an effective approach to generate assembly shellcodes and Python encoders through LLMs. The generation of shellcodes was also targeted by tools such as DualSC~\cite{Yang2022} and ExploitGen~\cite{Yang2023}, using a shallow transformer and a template augmented approach, respectively. \citeauthor{Chowdhary2023} in~\cite{Chowdhary2023} performed the Pentesting of web applications through LLMs by employing generative adversarial networks while~\citeauthor{Natella2024} in~\cite{Natella2024} introduced a novel dataset to train models in malicious Python generation. 

RedShell, while being significantly simpler than a LLM agent such as PentestGPT, still provides an efficient code generator to assist ethical hackers performing Pentest audits with more automation. In addition, our solution minimizes the amount of computational resources such as GPUs and VRAM required to train and evaluate AI-based code generators, following a much lighter approach than the generative adversarial networks employed by~\citeauthor{Chowdhary2023} in~\cite{Chowdhary2023}, and the pre-training strategies employed by~\citeauthor{Liguori2024} in~\cite{Liguori2024}. 

Furthermore, our specialized version of Qwen2.5-Coder outperformed in the most common output similarity metrics the current state-of-the-art solutions for malicious PowerShell generation, including~\cite{Liguori2024}, as demonstrated in Section~\ref{sec:tests}. Notably, RedShell was also able to outperform popular closed models such as ChatGPT 3.5 and DeepSeekChat-V3 in the semantic evaluation.

\section{Conclusions}
\label{sec:conclusions}

We demonstrated that RedShell is a competitive solution for assisting Pentesters in offensive PowerShell generation. Our tool generated malicious samples that were both syntactically valid and closely aligned with the reference snippets. We also built a ground truth dataset for malicious PowerShell, offering a strong knowledge base to train LLMs in offensive PowerShell generation.

A potential threat to the validity of RedShell lies in the absence of a functional evaluation of the generated snippets. Future work will address that by testing those snippets in controlled environments that mimic real-world Pentesting scenarios.

\printbibliography

\end{document}